\documentclass[final,3p,times,twocolumn]{elsarticle}

\usepackage{amssymb}
\usepackage{graphicx}
\usepackage{amsmath}

\newcommand{\bm}[1]{\mbox{\boldmath{$#1$}}}
\newcommand{\be}{\begin{eqnarray}}
\newcommand{\ee}{\end{eqnarray}}

\newcommand{\nn}{\nonumber}

\newcommand{\bo}{\boldsymbol}
\newcommand{\lb}{\label}

\newcounter{ichi}
\newcounter{ni}
\newcounter{san}
\newcounter{yon}
\newcounter{go}
\newcounter{roku}
\newcounter{nana}
\newcounter{hati}
\newcounter{kyu}
\journal{Physics Letters a}

\begin{document}

\begin{frontmatter}



\title{Extension of Nelson's Stochastic Quantization to Finite Temperature
Using Thermo Field Dynamics}

          
\author[KK]{K.~Kobayashi}
\ead{keita-x@fuji.waseda.jp}
\author[YY]{Y.~Yamanaka}
\ead{yamanaka@waseda.jp}
\address[KK]{Research Institute for Science and Engineering,
Waseda University, Tokyo 169-8555, Japan}
\address[YY]{Department of Electronic and Photonic Systems,
Waseda University, Tokyo 169-8555, Japan}
\begin{abstract}
We present an extension of Nelson's stochastic quantum mechanics
 to finite temperature.
 Utilizing the formulation of Thermo Field Dynamics (TFD), we can show that
 Ito's stochastic equations for tilde and non-tilde particle positions
 reproduce the TFD-type Schr\"odinger equation which is 
 equivalent to the Liouville-von Neumann equation.
 In our formalism, the drift terms in the Ito's stochastic equation have the
 temperature dependence and the thermal fluctuation is induced through the
 correlation of  the non-tilde and tilde particles.
 We show that our formalism satisfy the position-momentum uncertainty
 relation at finite temperature.
\end{abstract}

\begin{keyword}
Nelson stochastic mechanics \sep 
Thermo Field Dynamics 
\PACS
03.75.Lm \sep 03.75.Kk \sep 05.30.Jp
\end{keyword}

\end{frontmatter}


\section{Introduction}
\label{sec:level1}
The stochastic quantum mechanics was introduced by Nelson in 1966 \cite{Nelson}. 
There the motion of a particle in a microscopic world, governed by quantum theory,
is described by the hypothetical stochastic processes, as the Brownian motion. Thus the particle,
obeying the classical equation of motion, is affected by hypothetical random noise.
In Nelson's theory, one sets up the separate stochastic equations in forward and backward time 
directions, so that the temporal reversibility of microscopic motion is restored as a whole.
Actually, from the set of the forward and backward stochastic equations, and the Newton type
of the equation of motion, called  the Nelson-Newton equation, follows the reversible 
Schr\"odinger equation.

According to the Nelson's theory, we have an ensemble of an  infinite number of
classical trajectories (sample paths), and the average of any
 observable over them is equal to its expectation value,
given by the wave function in quantum mechanics. But the Nelson's theory offers us a wider
scope, i.e., we can calculate probability distributions and expectation values of
physical quantities which cannot be represented by operators, such as a time parameter. 
As an example, we refer to the tunneling time, arrival one and presence one of 
a particle in some region \cite{Imafuku1,Imafuku2,Hara1,Hara2}.

 After Nelson presented his theory, the extensions of the Nelson's approach
 were examined by several authors. A stochastic quantization based on
 stochastic action \cite{Yasue,Guerra} were introduced in 1980s. 
The generalization of Nelson's approach to interacting many particles were
 performed in Refs.~\cite{Loffrendo,KO}. 

The applications of the Nelson's stochastic quantum mechanics to dissipative
 system were investigated in Refs.~\cite{Yasue2,Ruggiero1,Ruggiero2,anharmonic,Misawa,Tbath}.
 For example, Ruggiero and Zannetti proposed a stochastic description of
the harmonic oscillator for the ground state in equilibrium with a thermal bath
 \cite{Ruggiero1,Ruggiero2}, and studied the anharmonic one later \cite{anharmonic}. 
Using a coherent state of the harmonic oscillator,
 they introduce the classical solution  and its quantum fluctuation which obeys 
the Nelson's stochastic quantum mechanics. 
Introducing the temperature dependent random noise to the classical equation,
 they constructed the diffusion process with thermal mixture for the harmonic oscillator. 
Their approach describes the dissipative process with quantum fluctuation
 and reduces to the Nelson's approach at zero temperature.
 But it is not
 clear whether their method fulfills the properties described 
by quantum statistical mechanics,
 such as position-momentum uncertainty
 relation at finite temperature \cite{Mann,Berman}.

The purpose of this paper is to extend the Nelson's stochastic quantum mechanics to
 thermal situation in framework of Thermo Field Dynamics (TFD) \cite{Umezawa}.
The average of a thermal mixed state is replaced by that of a pure state in
the TFD formulation at the price of doubling each degree of freedom: a new tilde 
degree of freedom is introduced to each original non-tilde one, and the state vector
space is a direct product of non-tilde and tilde state vector spaces.
In other words,  thermal fluctuations are realized as quantum fluctuations
through the correlation between non-tilde and tilde sectors.
Since in the Schr\"odinger picture of TFD the dynamical evolution is 
described by the Schr\"odinger equation with doubled degrees of freedom
 which is equivalent to the Liouville-von Neumann equation for the density matrix, 
we attempt to establish the Nelson's stochastic description to this TFD-type Schr\"odinger
equation.

This paper is organized as follows.
 In Sec.~2, we briefly review the Nelson's stochastic quantum mechanics
 and its equivalence to the Schr\"odinger equation.
 We introduce the TFD formalism and show its relation to the density
 matrix formalism in Sec.~3.
 Then we rewrite the TFD-type Schr\"odinger equation into the 
dynamical and kinematical equations
, which will be derived from the Nelson's  stochastic quantization.
 Section~4 is a main part of the paper, where we introduce the stochastic
equations and Nelson-Newton equations
 for the non-tilde and tide particles. The set of these equations 
turns out to be equivalent to
 the TFD-type Schr\"odinger equation. In Sec.~5, we apply our formalism to
 a harmonic oscillator and show the position-momentum uncertainty
 relation at finite temperature.
Section 6 is devoted to summary.
 
\section{Nelson's stochastic quantum mechanics}

In this section, we briefly review the Nelson's stochastic quantum mechanics
 and show its equivalence to the Schr\"odinger equation. Here, for simplicity,
 we consider the one-body Schr\"odinger equation:
\be
i\hbar\frac{\partial}{\partial t}\Psi(\bo{x},t)
=\left(-\frac{\hbar^{2}}{2m}\nabla^{2}+V(\bo{x},t)\right)\Psi(\bo{x},t)\,, \lb{eq:Scheq}
\ee
 with a mass $m$ and an external potential $V(\bo{x},t)$.
 When we rewrite the wave function $\Psi$ into the form $\Psi=e^{R+iS}$, where $R$ and $S$
are real functions, and define the current and osmotic velocities as
\be
&&\bo{v}=\frac{\hbar}{m}\nabla S\,,\\
&&\bo{u}=\frac{\hbar}{m}\nabla R\,,
\ee
the gradient of equation (\ref{eq:Scheq}) is transformed into the two equations, called
the dynamical and kinematical equations, respectively,
\be
&&\frac{\partial \bo{v}}{\partial t}=
\frac{\hbar}{2m}\nabla^{2}\bo{u}+
\bo{u}\cdot(\nabla\bo{u})-\bo{v}\cdot(\nabla\bo{v})
-\frac{1}{m}\nabla V \,,\nn\\
 \lb{eq:veq}\\
&&\frac{\partial \bo{u}}{\partial t}=
-\frac{\hbar}{2m}\nabla^{2}\bo{v}
-\nabla(\bo{u}\cdot\bo{v}) \lb{eq:ueq}\,.
\ee 

It can be shown as below that the Nelson's stochastic quantum mechanics leads to
equations (\ref{eq:veq}) and (\ref{eq:ueq}).

The first assumption of the Nelson's stochastic quantization is  the 
stochastic differential equations for the particle position $\bo{x}(t)$.
Since a single stochastic equation describes an irreversible process,
we set up a pair of stochastic equations for forward and backward
time evolutions separately, aiming at the reversible Schr\"odinger equation,
explicitly
\be
&&d\bo{x}(t)=
\bo{b}(\bo{x}(t),t)dt+\sqrt{\frac{\hbar}{m}}d\bo{W}\lb{eq:FSE}\,, \\
&&\left(d\bo{x}(t)=\bo{x}(t+dt)-
\bo{x}(t)\right)\nn\,,
\ee
for forward time evolution and 
\be
&&d\bo{x}(t)=\bo{b}_{*}(\bo{x}(t),t)dt+\sqrt{\frac{\hbar}{m}}d\bo{W}_{*}\,,\lb{eq:BSE} \\
&&\left(d\bo{x}(t)=\bo{x}(t)-\bo{x}(t-dt)\right)\,,\nn
\ee
for backward one ($dt>0$). Here $\bo{W}$ and $\bo{W}_{*}$ are
the standard Wiener processes
\be
E[dW_{i}(t)]=E[dW_{i*}(t)]=0\qquad i=x,y,z\,, \\
 E[dW_{i}(t)dW_{j}(t)]=E[dW_{*i}(t)dW_{*j}(t)]=\delta_{ij}dt\,,
\ee 
where $E[\cdots]$ means a sample average.

The second assumption is the Nelson-Newton equation of motion, defining the mean acceleration
 $\bo{a}(t)$ of random variables $\bo{x}(t)$ as follows:
\be
m\bo{a}(t)=-\nabla V \lb{eq:Newton}\,.
\ee
Here, the mean acceleration $\bo{a}(t)$ is defined as
\be
\bo{a}(t)=\frac{1}{2}(DD_{*}+D_{*}D)\bo{x}(t)\,.
\ee
with the mean forward time derivative $D$\,,
\be
\lefteqn{Df(t)=\lim_{dt\to 0+}E\left[\frac{f(\bo{x}(t+dt))
-f(\bo{x}(t))}{dt}\,\bigg{|}\,\bo{x}(t)\right]\,,}\nn\\
\ee
and the mean backward one $D_{*}$\,, 
\be
\lefteqn{D_{*}f(t)=\lim_{dt\to 0+}E\left[\frac{f(\bo{x}(t))-f(\bo{x}(t-dt))}{dt}
\,\bigg{|}\,\bo{x}(t)\right]\,,}\nn\\
\ee
 where $E[\cdots|\bo{x}(t)]$ means the conditional expectation.
From equations (\ref{eq:FSE}) and (\ref{eq:BSE}), the expression of the mean acceleration $\bo{a}(t)$
 is obtained as
\be
&&\bo{a}(t)=\frac{\partial}{\partial t}\frac{(D+D_{*})\bo{x}}{2}
+\frac{1}{2}(D_{*}\bo{x}\cdot \nabla)D\bo{x} \nn \\
&&+\frac{1}{2}(D\bo{x}\cdot \nabla)D_{*}\bo{x}-\frac{\hbar}{2m}\nabla^{2}
\frac{(D-D_{*})\bo{x}}{2} \lb{eq:Macce}\,.
\ee
Let us define the current and osmotic velocities as
\be
&&\bo{v}=\frac{1}{2}(\bo{b}+\bo{b}_{*})=\frac{1}{2}(D+D_{*})\bo{x}\,, \\
&&\bo{u}=\frac{1}{2}(\bo{b}-\bo{b}_{*})=\frac{1}{2}(D-D_{*})\bo{x}\,,
\ee
where we utilize the relations, $\bo{b}=D\bo{x}$ and $\bo{b}_{*}=D_{*}\bo{x}$. 
Then, substituting equation (\ref{eq:Macce}) into the Nelson-Newton equation of 
motion (\ref{eq:Newton}), we obtain the equation for the current velocity:
\be
\frac{\partial}{\partial t}\bo{v}=\frac{\hbar}{2m}\nabla^{2}\bo{u}
-(\bo{v}\cdot \nabla)\bo{v}+(\bo{u}\cdot \nabla)\bo{u}
-\frac{1}{m}\nabla V \,.
\lb{eq:NelsonCeq}
\ee

The stochastic processes of the random variable $\bo{x}(t)$ in equations (\ref{eq:FSE})
 and (\ref{eq:BSE}) are equivalently formulated by means of the distribution
function $P(\bo{x},t)$ which satisfy the Fokker-Planck equations,
\be
\frac{\partial}{\partial t}P=-\nabla\cdot(\bo{b}P)+\frac{\hbar}{2m}\nabla^{2}P
 \lb{eq:fFP}
\ee
for forward time  and
\be
\frac{\partial}{\partial t}P=-\nabla\cdot(\bo{b}_{*}P)-\frac{\hbar}{2m}\nabla^{2}P
 \lb{eq:bFP}
\ee
for backward time.  The sum of equations (\ref{eq:fFP}) and
(\ref{eq:bFP}) is the continuity equation
\be
\frac{\partial}{\partial t}P+\nabla\cdot\left(\bo{v}P\right)=0 \lb{eq:Aeq}\,,
\ee
while their difference yields the relation
\be
\nabla \cdot \left\{ \bo{u} P- \frac{\hbar}{2m}\nabla P\right\}=0 \lb{eq:Deq}\,.
\ee
Though the last equation implies $\bo{u} P- \frac{\hbar}{2m}\nabla P
= \nabla \times \bo{C}$ with an arbitrary vector function $\bm{C}(\bo{x},t)$,
it is shown \cite{Nelson2} that $\nabla \times \bo{C}=\bo{0}$, i.e.,
\be
\bo{u}=\frac{\hbar}{2m}\nabla\ln P \lb{eq:Seq}\,.
\ee
From equations (\ref{eq:Aeq}) and (\ref{eq:Seq}), one derives
\be
\frac{\partial}{\partial t}\bo{u}&=&\frac{\hbar}{2m}\frac{\partial}{\partial t}\nabla\ln P \nn \\
&=&-\frac{\hbar}{2m}\nabla (\nabla \cdot\bo{v})-\nabla(\bo{u}\cdot\bo{v})\,.
\ee
Assuming that $\bo{v}$ is given by a gradient of a scalar function, we 
have the equation for the osmotic velocity as
\be
\frac{\partial}{\partial t}\bo{u}
=-\frac{\hbar}{2m}\nabla^{2}\bo{v}-\nabla(\bo{u}\cdot\bo{v})\,, 
\lb{eq:NelsonOeq}
\ee
using $\bm {0}=
\nabla \times (\nabla \times \bo{v}) = \nabla (\nabla \cdot\bo{v})-\nabla^2 \bo{v}$.

Thus the equations of the current and osmotic velocities derived from Nelson's 
stochastic quantum mechanics, equations (\ref{eq:NelsonCeq}) and (\ref{eq:NelsonOeq}),
 are identical to equations (\ref{eq:veq}) and (\ref{eq:ueq}) originating from
 the Schr\"odinger equation.
\section{Thermo Field Dynamics}

In this section, we introduce a formulation of Thermo Field Dynamics (TFD)
in which the average over a density matrix is replaced with a pure state average in 
the doubled Hilbert space. To each original operator $A$, called a non-tilde operator, we 
introduce a new independent one ${\tilde A}$, called a tilde one.

The eigenvalue problems for the non-tilde and tilde Hamiltonians are
\be
&&H|u_{n}\rangle=E_{n}|u_{n}\rangle\, .\qquad |u_{n}\rangle\in\mathcal{H} \,,\\
&&\tilde{H}|\tilde{u}_{n}\rangle=E_{n}|\tilde{u}_{n}\rangle \, .\qquad
|\tilde{u}_{n}\rangle\in \tilde{\mathcal{H}}\,,
\ee
respectively. We construct a pure state vector $|\Phi\rangle$ from a superposition
of particular vectors $|u_{n},\tilde{u}_{n}\rangle\in\hat{\mathcal{H}}$ with a common $n$ 
in the doubled Hilbert 
space $\hat{\mathcal{H}}=\mathcal{H}\otimes\tilde{\mathcal{H}}$ with real coefficients $f_n$ as
\be
|\Phi\rangle&=&\sum_{n}f_{n}|u_{n},\tilde{u}_{n}\rangle\,.
\lb{eq:Phi}
\ee
Then the expectation value of any non-tilde operator $A$ for this $|\Phi\rangle$ becomes
\be
&&\langle\Phi|A|\Phi\rangle=\sum_{n} f_{n}^{2}\langle u_{n}|A|u_{n}\rangle
=Tr[\rho A] \lb{eq:Pav} \,, \\
&&\qquad\quad\rho=\sum_{n}  f_n^{2}|u_{n}\rangle\langle u_{n}|\,,
\ee 
where the ortho-normal condition $\langle u_n| u_m \rangle=
\langle {\tilde u}_n| {\tilde u}_m \rangle=\delta_{nm}$ has been used.
This way the expectation value for the pure state vector $|\Phi\rangle$, which is 
called a thermal vacuum, 
is equivalent to the trace average over the density matrix $\rho$.
If we set
$f_{n}^{({\rm eq})}={e^{-\frac{\beta E_{n}}{2}}}/{(\sum_{n}e^{-\beta E_{n}})^{1/2}}$\, ,
where $\beta$ is the inverse temperature, equation (\ref{eq:Pav}) is reduced 
to the trace average in thermal equilibrium.
We may say that the TFD formalism represents the thermal fluctuation 
through the quantum correlation  between non-tilde and tilde particles.

In TFD, the total Hamiltonian of the system is given by 
\be
\hat{H}=H-\tilde{H}\,,
\lb{eq:hatH}
\ee
 and the Heisenberg operators $A(t)$ and $\tilde{A}(t)$ obey the Heisenberg
 equations
\be
&&i\hbar\frac{d}{d t}A=[A,\hat{H}]=[A,H]\,, \\
&&i\hbar\frac{d}{d t}\tilde{A}=[\tilde{A},\hat{H}]=-[\tilde{A},\tilde{H}]\,. 
\ee
Note that all the canonical commutation relations between non-tilde and tilde operators vanish.
The minus sign in front of ${\tilde H}$ in equation (\ref{eq:hatH})
 is crucial, which is required from the stability of the thermal vacuum $|\Phi \rangle$ 
in equation (\ref{eq:Phi}) including the thermal equilibrium  \cite{Umezawa}.
It also implies that the tilde system corresponds to 
the time reversed one of the non-tilde particle.

The exchange operations of non-tilde and tilde operators are summarized in the 
tilde conjugation rules \cite{Umezawa}
in TFD:
\be
&&(c_1A+ c_2B)^{\thicksim}= c_1^\ast {\tilde A}+c_2^\ast {\tilde B}\\
&& ({\tilde A})^{\thicksim}=A \\
&& (A^\dagger)^{\thicksim}={\tilde A}^\dagger \\
&& (AB)^{\thicksim}= {\tilde A}{\tilde B} \, .
\lb{eq:TildeConOp}
\ee
where $c_{1}$ and $c_{2}$ are complex number.
We note that the thermal vacuum is invariant under the tilde conjugation:
\be
|\Phi \rangle^{\thicksim}= |\Phi \rangle \, , \quad
\langle \Phi |^{\thicksim}=\langle \Phi | \, ,
\lb{eq:TildeConTV}
\ee
where $(c|u_{n},\tilde{u}_{m}\rangle)^{\thicksim}=c^{*}|u_{m},\tilde{u}_{n}\rangle$\,. 

The time evolution of any state vector $|\Psi(t)\rangle$ in the Schr\"odinger picture
is governed by the TFD-type Schr\"odinger equation:
\be
i\hbar\frac{d}{d t}|\Psi(t)\rangle=(H-\tilde{H})|\Psi(t)\rangle
={\hat H}|\Psi(t)\rangle
\,.\lb{eq:TFDsch0}
\ee
For the thermal vacuum $|\Phi \rangle$, this equation reads as
\be
0= {\hat H} |\Phi\rangle \, ,
\ee
implying that the thermal vacuum is a stationary state with zero eigenvalue.
We require that the state vector $|\Psi(t)\rangle$ describing thermal situation
should satisfy the tilde invariance
\be
(|\Psi(t)\rangle)^{\thicksim}=|\Psi(t)\rangle \lb{eq:tildeinvar}\, .
\ee
The reason for this will be clear soon, in connection to the equivalence
 to the density matrix formalism.

As an example of using equation (\ref{eq:TFDsch0}), one can imagine the case in which 
an isolated system is
 initially in a stationary thermal vacuum in equation (\ref{eq:Phi})
and a time-dependent external potential is switched on at $t=0$\,.
 Then the total Hamiltonian is time-dependent, $\hat{H}(t)=H(t)-\tilde{H}(t)$ and
 the state vector is given by
\be
&&|\Psi(t)\rangle=\sum_{n}f_{n}|u_{n}(t),\tilde{u}_{n}(t)\rangle
\,,
\ee
where $|u_{n}(t)\rangle$ and $|\tilde{u}_{n}(t)\rangle$
 obey the following equations
\be
&&i\hbar\frac{d}{d t}|u_{n}(t)\rangle=H(t)|u_{n}(t)\rangle \,, \\
&&i\hbar\frac{d}{d t}|\tilde{u}_{n}(t)\rangle=-\tilde{H}(t)|\tilde{u}_{n}(t)\rangle \,, 
\ee
under the initial conditions: $|u_{n}(t=0)\rangle=|u_{n}\rangle$
 and $|\tilde{u}_{n}(t=0)\rangle=|\tilde{u}_{n}\rangle$.
Note that the  tilde invariance of the state vector $|\Psi(t)\rangle$ is
 preserved under the time evolution.

The TFD formalism has a one-to-one correspondence to the super-operator formalism 
for density matrix \cite{Sch,Arimitsu1}.
In the Schr\"odinger picture of quantum statistical mechanics, 
the density matrix $\rho(t)$ obeys the
 Liouville-von Neumann equation
\be
i\hbar\frac{d}{d t}\rho(t)=\mathcal{L}\rho(t) \,,
\ee
where $\mathcal{L}$ is the Liouville operator:
\be
\mathcal{L}\cdots =[H,\cdots]\,.
\ee
In the super-operator formalism, the density matrix is treated as a vector
 in a vector space on which an operator acts in two ways:
 an operation from the left-hand side of $\rho$ as $A\rho$ , and the other one
 from the right-hand side of $\rho$ as $\rho A$. We may rewrite right side operation 
as ${\tilde{A}}^\dagger\rho=\rho A$.
This is the way how every operator is doubled and
 the operator $A$ accompanies its tilde conjugation $\tilde{A}$.
 Furthermore, rewriting the Liouville operator as $\mathcal{L}\rho=(H-\tilde{H})\rho$,
 we can recognize the form of the TFD Hamiltonian: $H-\tilde{H}$. 
The state vector $|\Psi\rangle$ in the doubled Hilbert space corresponds to 
a density matrix $\rho$ and its tilde invariance in equation (\ref{eq:tildeinvar}) corresponds
 to the hermiticity of the density matrix $\rho^{\dagger}=\rho$.
 The correspondence between the super-operator formalism and
TFD  is as follows, 
\be
\rho &\leftrightarrow& |\Psi\rangle \,,\\
A\rho &\leftrightarrow& A|\Psi\rangle \,,\\
\rho A^{\dagger}&\leftrightarrow& \tilde{A}|\Psi\rangle \,,\\
\mathcal{L} &\leftrightarrow& \hat{H}=H-\tilde{H} \,,
\ee
and
\be
&&i\hbar\frac{d}{d t}\rho(t)=\mathcal{L}\rho(t) \nn\\
&\leftrightarrow&
i\hbar\frac{d}{d t}|\Psi(t)\rangle=(H-\tilde{H})|\Psi(t)\rangle\,.
\ee

Finally, we rewrite the TFD-type Schr\"odinger equation into the from of 
the dynamical and  kinematical equations.
In the coordinate representation ($\displaystyle \Psi(\bo{x},\tilde{\bo{x}},t)= 
\langle\bo{x},\tilde{\bo{x}}|\Psi(t)\rangle$), 
the TFD-type Schr\"odinger equation is given by
\be
\lefteqn{i\hbar\frac{\partial}{\partial t}\Psi(\bo{x},\tilde{\bo{x}},t)
=\left[H(\nabla,\bo{x},t)-H(\tilde{\nabla},\tilde{\bo{x}},t)\right]\Psi(\bo{x},\tilde{\bo{x}},t) \,,}\nn\\
\ee
with
\be
&&H(\nabla,\bo{x},t)=
-\frac{\hbar^{2}}{2m}\nabla^{2}+V(\bo{x},t)
 \lb{eq:TFDScheq} \,, 
\ee
and the tilde invariance
\be
\Psi(\bo{x},{\tilde{\bo{x}}},t)=
\left(\Psi(\bo{x},{\tilde{\bo{x}}},t)\right)^{\thicksim}
=\Psi^{*}({\tilde{\bo{x}}},\bo{x},t)\,.
\ee
Now, let us put the wave function $\Psi=e^{R+iS}$ and
 introduce the current and osmotic velocities as
\be
&&\bo{v}=\frac{\hbar}{m}\nabla S\,,\quad
 {\tilde {\bo{v}}}=-\frac{\hbar}{m}\tilde{\nabla} S\,,\\
&&\bo{u}=\frac{\hbar}{m}\nabla R\,, \quad
{\tilde{\bo{u}}}=\frac{\hbar}{m}\tilde{\nabla} R\,.
\ee
Then  equation (\ref{eq:TFDScheq}) is transformed into the dynamical and kinematical equations
\be
\frac{\partial \bo{v}}{\partial t}&=&
\frac{\hbar}{2m}(\nabla^{2}-\tilde{\nabla}^{2})\bo{u}+
(\bo{u}\cdot\nabla-{\tilde{\bo{u}}}\cdot\tilde{\nabla})\bo{u}
 \nn \\
&&-(\bo{v}\cdot\nabla+{\tilde{\bo{v}}}\cdot\tilde{\nabla})\bo{v}
-\frac{1}{m}\nabla (V-\tilde{V}) \lb{eq:TFDveq}\,,\\
\frac{\partial \bo{u}}{\partial t}&=&
-\frac{\hbar}{2m}(\nabla^{2}-\tilde{\nabla}^{2})\bo{v}
-\nabla(\bo{u}\cdot\bo{v}+{\tilde{\bo{u}}}\cdot {\tilde{\bo{v}}}) \lb{eq:TFDueq}\, ,
\ee 
and their tilde conjugates. The notation $\tilde V$ stands for $V({\tilde{\bo{x}}},t)$.

\section{Nelson's Stochastic Quantization in Thermo Field Dynamics}

In this section, we extend the Nelson's stochastic quantization to 
systems described by TFD.
Our aim is to drive the dynamical and kinematical equations (\ref{eq:TFDveq}) and 
(\ref{eq:TFDueq}), equivalent
 to the Liouville-von Neumann equation, from the  stochastic differential equations.
 
A naive extension of non-tilde system to non-tilde and tilde one is unsuccessful,
since then the total Hamiltonian would be not $H-{\tilde H}$ but $H+{\tilde H}$\, .
In order to accommodate the minus sign in front of ${\tilde H}$, we classify the four
stochastic differential equations into the following two groups:
one is a pair of those for $\bo{x}(t)$ in forward evolution and ${\tilde {\bo{x}}}(t)$
in backward one
\be
d\bo{x}(t)&=&\bo{b}(\bo{x}(t),\tilde{\bo{x}}(t),t)d t+\sqrt{\frac{\hbar}{m}}d\bo{W} \,, \\
&&\left(d\bo{x}(t)=\bo{x}(t+dt)-\bo{x}(t) \right)\,,\nn \\
d\tilde{\bo{x}}(t)&=&{\tilde{\bo{b}}}_{*}(\bo{x}(t),\tilde{\bo{x}}(t),t)d t+
\sqrt{\frac{\hbar}{m}}d\tilde{\bo{W}}_{*} \,, \\
&&\left(d{\tilde {\bo{x}}}(t)
={\tilde{\bo{x}}}(t)-{\tilde{\bo{x}}}(t-dt)\right)\nn \,,
\ee
and the other is a pair of those for $\bo{x}(t)$ in  backward evolution
 and ${\tilde {\bo{x}}}(t)$
in forward one
\be
d\bo{x}(t)&=&\bo{b}_{*}(\bo{x}(t),\tilde{\bo{x}}(t),t)d t+\sqrt{\frac{\hbar}{m}}d\bo{W}_{*}\,, \\
&&\left(d\bo{x}(t)=\bo{x}(t)-\bo{x}(t-dt)\right) \,, \nn\\
d\tilde{\bo{x}}(t)&=&\bo{\tilde{b}}(\bo{x}(t),\tilde{\bo{x}}(t),t)d t+
\sqrt{\frac{\hbar}{m}}d\tilde{\bo{W}} \,, \\
&& \left(d\bo{\tilde{x}}(t)=\bo{\tilde{x}}(t+dt)-\bo{\tilde{x}}(t)\right)\,.\nn
\ee
Here $d\bo{W}\, ,d\tilde{\bo{W}}_{*}\, ,d\bo{W}_{*} \, , d\tilde{\bo{W}}$ are independent
Wiener processes. The above grouping is suggested by the fact that  
the equation of the tilde particle is a time reversed one of the non-tilde particle 
in TFD.

A crucial step in our present formulation is to introduce 
new mean time derivatives by
\be
&&{\bar D}f(\bo{x}(t),\tilde{\bo{x}}(t),t) \nn \\
&&=\lim_{dt\to 0+}E\left[\frac{1}{dt}\left\{f(\bo{x}(t+dt),\tilde{\bo{x}}(t),t+dt)\right.\right.\nn\\
&&\left.\left.\qquad-f(\bo{x}(t),\tilde{\bo{x}}(t-dt),t)\right\}\bigg{|}\,\bo{x}(t),\tilde{\bo{x}}(t)\right]\,,\nn\\
\\
&&{\bar D}_{*}f(\bo{x}(t),\tilde{\bo{x}}(t),t) \nn \\
&&=\lim_{dt\to 0+}E\left[\frac{1}{dt}\left\{f(\bo{x}(t),\tilde{\bo{x}}(t+dt),t)\right.\right.\nn\\
&&\left.\left.\qquad-f(\bo{x}(t-dt),\tilde{\bo{x}}(t),t-dt)\right\}\bigg{|}\,\bo{x}(t),\tilde{\bo{x}}(t)\right] \,,\nn\\
\ee
The mean derivatives ${\bar D}$ and ${\bar D}_{*}$ are hybrids of the original $D$ and $D_{*}$,
and
give the formulas for $f(\bo{x}(t),\tilde{\bo{x}}(t),t)$\,,
\be
{\bar D}f&=&\frac{\partial f}{\partial t}+
\bo{b}\cdot(\nabla f)+
\tilde{\bo{b}}_{*}\cdot(\tilde{\nabla} f)\nn\\
&&+\frac{\hbar}{2m}\nabla^{2}f-
\frac{\hbar}{2m}\tilde{\nabla}^{2}f\,, \\
{\bar D}_{*}f&=&\frac{\partial f}{\partial t}+
\bo{b}_{*}\cdot(\nabla f)+
\tilde{\bo{b}}\cdot(\tilde{\nabla} f)\nn\\
&&+\frac{\hbar}{2m}\nabla^{2}f-
\frac{\hbar}{2m}\tilde{\nabla}^{2}f\,.
\ee
Then, utilizing the relations for any function $f(\bo{x}(t),\tilde{\bo{x}}(t))$\,,
\be
\frac{d}{d t}E[f]=E[{\bar D}f]=E[{\bar D}_{*}f]\,,
\ee
we can obtain the Fokker-Planck equations for the distribution
 $P(\bo{x},\tilde{\bo{x}},t)$\,,
\be
\frac{\partial P}{\partial t}&=&-
\nabla\cdot(\bo{b}P)-\tilde{\nabla}\cdot({\tilde{\bo{b}}}_{*}P)\nn\\
&&+\frac{1}{2}\frac{\hbar}{m}\left(\nabla^{2}-\tilde{\nabla}^{2}\right)P \lb{eq:tfdFFPequ}\,, \\
\frac{\partial P}{\partial t}&=&-
\nabla\cdot(\bo{b}_{*}P)-\tilde{\nabla}\cdot(\tilde{\bo{b}}P)\nn\\
&&-\frac{1}{2}\frac{\hbar}{m}\left(\nabla^{2}-\tilde{\nabla}^{2}\right)P 
\lb{eq:tfdBFPequ}\,.
\ee
Here, we define the current and osmotic velocities of the non-tilde and tilde particles as
\be
&&\bo{u}=\frac{1}{2}(\bo{b}-\bo{b}_{*})\,, \quad
\bo{v}=\frac{1}{2}(\bo{b}+\bo{b}_{*})\,, \\
&&\tilde{\bo{u}}=\frac{1}{2}(\tilde{\bo{b}}-\tilde{\bo{b}}_{*})\,, \quad
\tilde{\bo{v}}=\frac{1}{2}(\tilde{\bo{b}}+\tilde{\bo{b}}_{*}) \,,
\ee
respectively.
The difference of equations (\ref{eq:tfdFFPequ}) and (\ref{eq:tfdBFPequ}) gives us the relation:
\be
\nabla\cdot\left(\bo{u}P-\frac{\hbar}{2m}\nabla P\right)
-\tilde{\nabla}\cdot\left((\tilde{\bo{u}}P)-\frac{\hbar}{2m}\tilde{\nabla}P\right)=0\,.
\ee
The argument to derive equation (\ref{eq:Seq}) applies here as well, so we have
\be
\bo{u}=\frac{\hbar}{2m}\nabla\ln P\,,\qquad 
\bo{\tilde{u}}=\frac{\hbar}{2m}\tilde{\nabla}\ln P \lb{eq:A1}\,.
\ee
The sum of equations (\ref{eq:tfdFFPequ})  and (\ref{eq:tfdBFPequ}) 
is the continuity equation,
\be
\frac{\partial }{\partial t}P+\nabla\cdot(\bo{v}P)
+\tilde{\nabla}\cdot(\tilde{\bo{v}}P)=0 \lb{eq:tfdCeq}\,.
\ee
From equations (\ref{eq:A1}) and (\ref{eq:tfdCeq}) follows the TFD-type 
kinematical equation (\ref{eq:TFDueq}) as
\be
\frac{\partial}{\partial t}\bo{u}&=&
\frac{\hbar}{2m}\nabla\left(\frac{1}{P}\frac{\partial P}{\partial t}\right)\nn \\
&=&\frac{\hbar}{2m}\nabla\left\{-\frac{1}{P}
\nabla\cdot(\bo{v}P)-\frac{1}{P}\tilde{\nabla}\cdot(\tilde{\bo{v}}P)
\right\} \nn \\
&=&\nabla\left\{-\frac{\hbar}{2m}\nabla\cdot\bo{v}
-\frac{\hbar}{2m}\tilde{\nabla}\cdot\tilde{\bo{v}}
-\bo{v}\cdot\bo{u}-\bo{\tilde{v}}\cdot\bo{\tilde{u}}
\right\} \nn \\
&=&-\frac{\hbar}{2m}\left( \nabla^{2}-\tilde{\nabla}^{2}\right) \bo{v}
-\nabla(\bo{u}\cdot\bo{v}+\tilde{\bo{u}}\cdot\tilde{\bo{v}})\,.
\ee
In the last equality we have assumed that $\bo{v}$ and $\tilde{\bo{v}}$ are gradients
of a single function $g(\bo{x},\tilde{\bo{x}},t)$ as
$
\bo{v}= \nabla g
$ and 
$ -\tilde{\bo{v}}={\tilde \nabla} g
$\,,
so that 
$
\nabla(\nabla\cdot \bo{v}) = \nabla^2 \bo{v}$ and 
$\nabla(\tilde \nabla\cdot \tilde {\bo{v}}) =-\nabla{\tilde \nabla}^2g
= -{\tilde \nabla}^2\bo{v}$\,.
Next, we derive an expression of the mean acceleration $\bo{a}(t)$.
Its natural candidate would be an average of 
the quantities ${\bar D}_{*}{\bar D}\bo{x}(t)$ and ${\bar D}{\bar D}_{*}\bo{x}(t)$,
each of which can be computed as
\be
{\bar D}_{*}{\bar D}\bo{x}&=&\frac{\partial \bo{b}}{\partial t}+
(\bo{b}_{*}\cdot\nabla) \bo{b}+
(\bo{\tilde{b}}\cdot \tilde{\nabla})\bo{b} \nn \\
&&-\frac{\hbar}{2m}(\nabla^{2}-\tilde{\nabla}^{2} )\bo{b}
\,, \\
{\bar D}{\bar D}_{*}\bo{x}&=&\frac{\partial \bo{b}_{*}}{\partial t}+
(\bo{b}\cdot\nabla) \bo{b}_{*}+
({\tilde{\bo{b}}}_{*}\cdot \tilde{\nabla})\bo{b}_{*}\nn\\
&&+\frac{\hbar}{2m}(\nabla^{2}-\tilde{\nabla}^{2})\bo{b}_{*}
\,, 
\ee
and their tilde conjugates. 
Then, the mean accelerations $\bo{a}$ and $\tilde{\bo{a}}$ are given by
\be
\bo{a}&=&\frac{\partial \bo{v}}{\partial t}
-
\frac{\hbar}{2m}\left(\nabla^{2}-
\tilde{\nabla}^{2}
\right)\bo{u}\nn \\
&&+(\bo{v}\cdot\nabla+\tilde{\bo{v}}\cdot\tilde{\nabla})\bo{v}-
(\bo{u}\cdot\nabla-\tilde{\bo{u}}\cdot\tilde{\nabla})\bo{u}
\,, \lb{eq:tfdaeq}
\ee
and its tilde conjugate. We require the following Nelson-Newton equations,
\be
&&m\bo{a}=-\nabla (V-\tilde V) \, , \\
&&m{\tilde{ \bo{a}}}= {\tilde \nabla}(V-\tilde V)\, .
\ee  
Substituting equation (\ref{eq:tfdaeq}) into the former equation, 
we obtain the TFD-type dynamical equation: 
\be
\frac{\partial \bo{v}}{\partial t}&=&\frac{\hbar}{2m}\left(
\nabla^{2}-
\tilde{\nabla}^{2}
\right)\bo{u}+(\bo{u}\cdot\nabla-\tilde{\bo{u}}\cdot\tilde{\nabla})\bo{u}\nn \\
&&-
(\bo{v}\cdot\nabla+\tilde{\bo{v}}\cdot\tilde{\nabla})\bo{v}-
\frac{1}{m}\nabla (V-{\tilde V})\,.
\ee
Consequently, if any solution of the TFD-type Schr\"odinger equation 
$\Psi(\bo{x},\tilde{\bo{x}},t) =e^{R+iS}$ is given,
 the drift terms in the stochastic equations are given by the gradients of
$R$ and $S$ as
\be
&&\bo{b}=\frac{\hbar}{m}\nabla(R+S) \,,\\
&&\bo{b}_{*}=-\frac{\hbar}{m}\nabla(R-S) \,,\\
&&\bo{\tilde{b}}=\frac{\hbar}{m}\tilde{\nabla}(R-S)\,, \\
&&\bo{\tilde{b}}_{*}=-\frac{\hbar}{m}\tilde{\nabla}(R+S) \,.
\ee

Thus we have the Nelson's stochastic quantum mechanics, equivalent to the TFD-type
Schr\"odinger equation.

\section{Application to Harmonic Oscillator and Generalized Uncertainty Relation}

\begin{figure}[h]
\begin{center}
\includegraphics[width=1.00\linewidth]{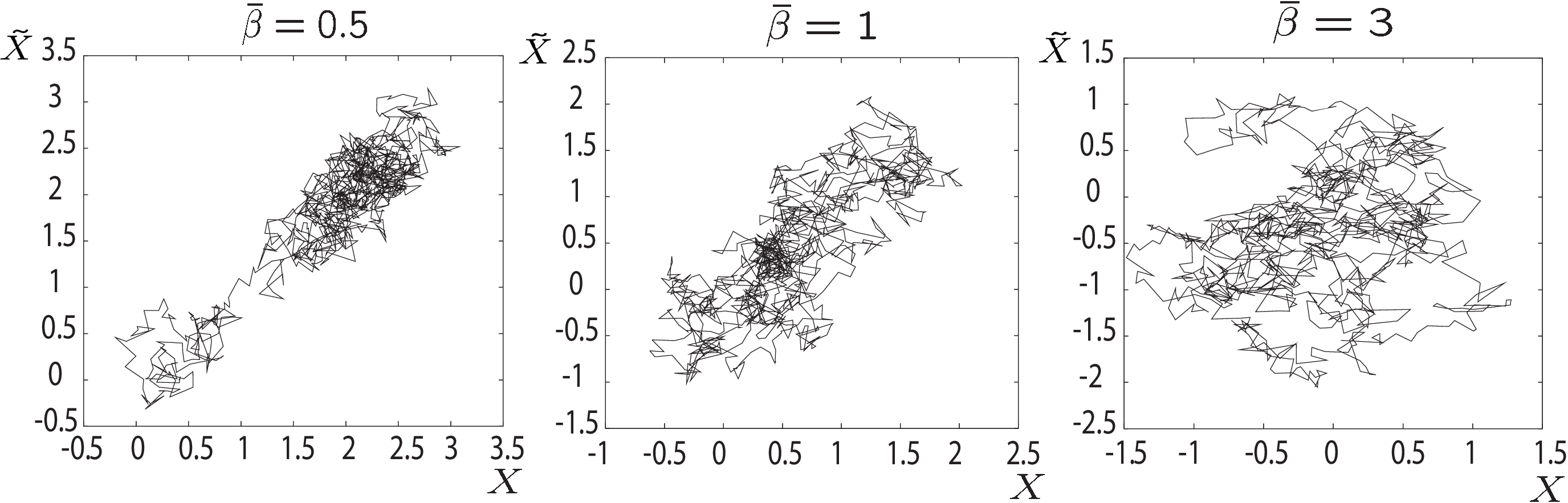}
\end{center}
\caption{\footnotesize{ 
The typical sample path on $(X,\tilde{X})$ plane for $\bar{\beta}=0.5$, $\bar{\beta}=1$ and $\bar{\beta}=3$ with $\bar{\beta}=\hbar\omega\beta$ and $X=\sqrt{\frac{m\omega}{\hbar}}x$.
}}
\label{Path}
\end{figure}

\begin{figure}[h]
\begin{center}
\includegraphics[width=0.90\linewidth]{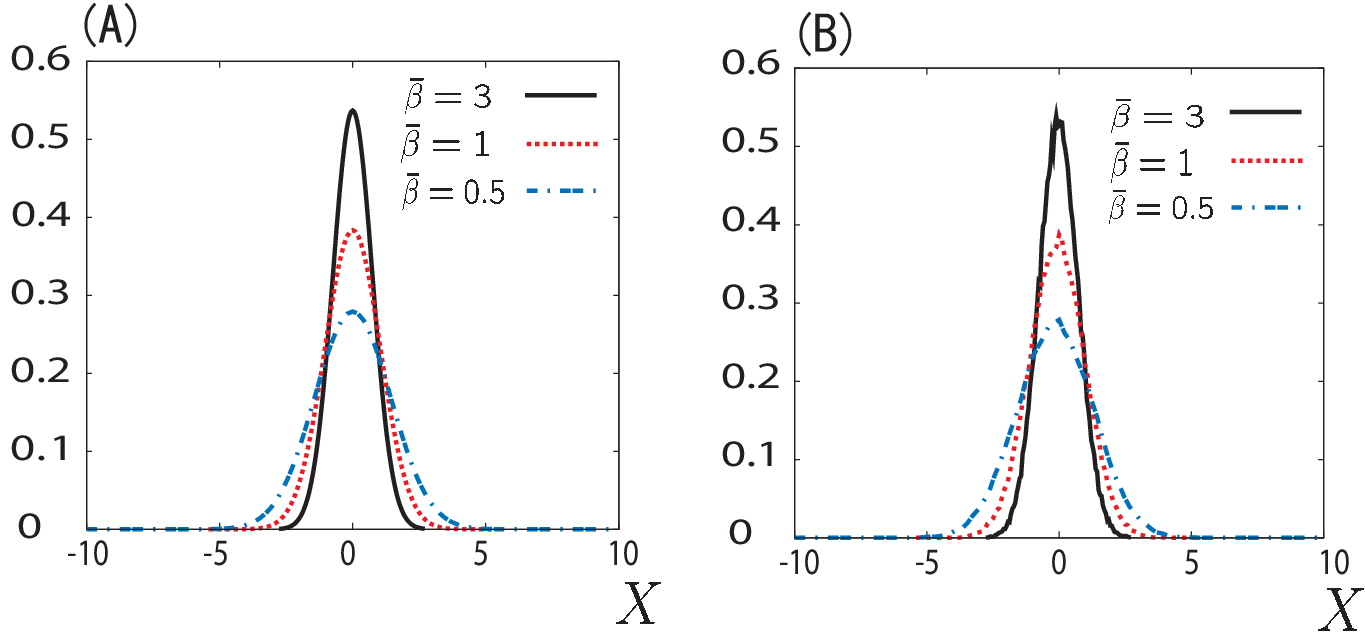}
\end{center}
\caption{\footnotesize{ 
Figure (A): The analytic solutions of the probability distribution
 for $\bar{\beta}=3$, $\bar{\beta}=1$ and $\bar{\beta}=0.5$
 with $\bar{\beta}=\hbar\omega\beta$ and $X=\sqrt{\frac{m\omega}{\hbar}}x$.
Figure (B): The probability distribution which is obtained numerically
 from $10^5$ sample paths of the non-tilde and tilde stochastic equations.
}}
\label{His}
\end{figure}

We apply the Nelson's stochastic quantum mechanics at finite temperature
in the previous section 
to a particle in one dimensional harmonic oscillator.
The TFD-type Schr\"odinger equation is
\be
&&i\hbar\frac{\partial}{\partial t}\Psi(x,\tilde{x},t)
=\left(H-\tilde{H}
\right)\Psi(x,\tilde{x},t)\,, \\
&&H=\left(
-\frac{\hbar^{2}}{2m}\frac{\partial^{2}}{\partial x^{2}}
+\frac{\omega^{2} m}{2}x^{2}\right)\,.
\ee
In thermal equilibrium, the wave function $\Psi$ is given by
\be
\lefteqn{\Psi_{{\rm eq}}(x,\tilde{x})=\sum_{n}\frac{e^{-\frac{\beta \hbar\omega n}{2}}}{Z(\beta)^{1/2}}u_{n}(x)u_{n}^{*}(\tilde{x}) \,,}\\
\lefteqn{\left(
-\frac{\hbar^{2}}{2m}\frac{\partial^{2}}{\partial x^{2}}
+\frac{\omega^{2} m}{2}x^{2}
\right)u_{n}(x)=\hbar\omega\left(n+\frac{1}{2}\right)u_{n}(x)\,,}\nn\\
\ee 
with $n=0,1,2,\cdots$. Using the Feynman kernel,
\be 
\psi(x_1,t_1)= \int\! K(x_1,t_1; x_2, t_2) \psi(x_2, t_2)\, dx_2 \,,
\ee
\be
&&K(x_{1},t_{1};x_{2},t_{2})=
\sqrt{\frac{m\omega}{2\pi \hbar \sin\omega(t_{1}-t_{2})}}e^{iS_{{\rm cl}}/\hbar}\,,\nn \\
\\
&&S_{{\rm cl}}=\frac{m\omega}{2\sin \omega(t_{1}-t_{2})} \nn\\
&&\qquad\times[(x_{1}^{2}+x_{2}^{2})\cos\omega(t_{1}-t_{2})-2x_{1}x_{2}]\,,
\ee
we obtain the analytic solution of the TFD-type Schr\"odinger equation
 in thermal equilibrium as
\be
\lefteqn{\Psi_{{\rm eq}}(x,\tilde{x})=\frac{1}{Z(\beta)^{1/2}}K(x,-i\hbar\beta/2;\tilde{x},0)\,,} \\
\lefteqn{Z(\beta)=\frac{1}{2\sinh(\beta\hbar\omega/2)}\,,} \\
\lefteqn{K(x,\tilde{x};-i\hbar\beta/2,0)=
\sqrt{\frac{m\omega}{2\pi\hbar\sinh(\beta\hbar\omega/2)}}e^{R_{{\rm eq}}}\,,} \\
\lefteqn{R_{{\rm eq}}=-\frac{m\omega}{\hbar}\frac{(x^{2}+\tilde{x}^{2})\cosh(\beta\hbar\omega/2)-2x\tilde{x}}{2\sinh(\beta\hbar\omega/2)}\,.}
\ee
Then, the drift terms in the stochastic equations are calculated as
\be
&&b_{{\rm eq}}(x(t),\tilde{x}(t))=\nn\\
&&-\omega\left(x(t)\frac{\cosh(\beta\hbar\omega/2)}{\sinh(\beta\hbar\omega/2)}-\tilde{x}(t)\frac{1}{\sinh(\beta\hbar\omega/2)}\right)\,, \nn\\
\\
&&\tilde{b}_{{\rm eq}*}(x(t),\tilde{x}(t))=\nn\\
&&\omega\left(\tilde{x}(t)\frac{\cosh(\beta\hbar\omega/2)}{\sinh(\beta\hbar\omega/2)}-x(t)\frac{1}{\sinh(\beta\hbar\omega/2)}\right)\,.\nn\\
\ee
Note that the drift term $b_{{\rm eq}}(x(t),\tilde{x}(t))$ has the temperature dependence because of the temperature dependence of the wave function $\Psi_{{\rm eq}}(x,\tilde{x})$.
In the limit of zero temperature $\beta\to \infty$,
 the drift term $b_{{\rm eq}}(x(t),\tilde{x}(t))$ is reduced to $b_{{\rm eq}}(x(t),\tilde{x}(t))\to-\omega x(t)$, that is, the correlation between the non-tilde and tilde particles disappears. 
This way, our formalism reproduces the original Nelson's stochastic quantum mechanics at zero temperature.
The numerical results of the non-tilde and tilde stochastic equations
 are shown in figures \ref{Path} and \ref{His}.  Figure 1 shows stronger correlations
 between $X$ and ${\tilde X}$ for higher temperature, namely larger thermal fluctuations.
One can see that the probability distribution from numerical calculations 
of sample paths fits the analytic one well.

Finally, we show the position-momentum uncertainty relation 
in stochastic mechanics \cite{FMS,FMS2}.
Taking the sample average of equation (\ref{eq:A1}), we obtain the relation,
 $E[\bo{p}]=E[\bo{p}_{*}]$,
where we have introduced the notations of $\bo{p}=m\bo{b}$ and
 $\bo{p}_{*}=m\bo{b}_{*}$. On the other hand, multiplying  equation (\ref{eq:A1}) by $x_{i}$,
we drive the relation, $E[x_{i}p_{j}-x_{i}p_{*j}]=-\hbar\delta_{ij}$. Utilizing the above relations, we can evaluate the quantity:
\be
\lefteqn{
E\left[(x_{i}-E[x_{i}])\left(\frac{p_{i}-p_{*i}}{2}-E\left[\frac{p_{i}-p_{*i}}{2}\right]\right)\right]=-\frac{\hbar}{2} \,.}\lb{eq:var1}\nn\\
\ee
The Schwarz's inequality is applied to equation (\ref{eq:var1}), then 
the position-momentum uncertainty relation in stochastic mechanics is given by
\be
\sqrt{{\rm Var[x_{i}]}}\sqrt{{\rm Var}[(p_{i}-p_{*i})/2]}
\ge \frac{\hbar}{2} \,.
\ee

The position and momentum uncertainty at finite temperature
for harmonic oscillator is checked explicitly. The variances
 of $x_{i}$ and $(p_{i}-p_{*i})/2$ are
\be
{\rm Var[x_{i}]}=\frac{\hbar}{2m\omega}\frac{1}{\tanh\left(\frac{\beta\hbar\omega}{2}\right)}\,, \\
{\rm Var}[(p_{i}-p_{*i})/2]=\frac{m\hbar\omega}{2}\frac{1}{\tanh\left(\frac{\beta\hbar\omega}{2}\right)}\,,
\ee
and the position and momentum uncertainty is given by
\be
&&\sqrt{{\rm Var[x_{i}]}}\sqrt{{\rm Var}[(p_{i}-p_{*i})/2]}=
\frac{\hbar}{2}+\hbar n \,, \\
&&n=\frac{1}{e^{\beta\hbar\omega}-1} \,. 
\ee
Note that the position and momentum uncertainty depend not only on 
 Planck constant $\hbar$ but also on the temperature of the system.
This result is consistent with the position-momentum uncertainty relation
 derived by thermal Bogoliubov transformation \cite{Mann,Berman}.

\section{Summary}
In this paper, we have extended the Nelson's stochastic quantum mechanics to
 thermal situation in framework of Thermo Field Dynamics (TFD) \cite{Umezawa}.
We set the four stochastic equations: the 
two are for the non-tilde particle in forward and backward
times, and the other two are for the tilde particle in forward and backward
times. The essence of our successful formulation is to make the pairs, i.e.,
 the pair of
 non-tilde operator in forward time
and tilde one in backward one, and the pair of non-tilde operator in backward time and tilde one
in forward one, taking into
 account that the time-evolution of the tilde system corresponds to 
the time reversed one of the non-tilde one. We have shown that the four stochastic
 and two Nelson-Newton equations reproduce the corresponding TFD-type Schr\"odinger equation, which is equivalent to the Liouville-von Neumann equation.
 In our formalism, the drift terms in the stochastic equations have the
 temperature dependence and the thermal fluctuation is induced through the
 quantum correlation between the non-tilde and tilde systems.
 In the limit of zero temperature, the temperature dependence in the drift terms disappears
 and our formulation reduces to the original Nelson quantum mechanics at zero temperature.
 Note that our formulation can be easily extended to many-body systems including
 systems of identical particles \cite{Loffrendo,KO}.
 In application of our theory to harmonic oscillator, we have analyzed 
the position-momentum uncertainty  relation at finite temperature.
Our combined formulation of the Nelson's stochastic quantum mechanics and TFD 
gives us a new insight into quantum theory in thermal situation.

Finally we comment on another view about the thermal and quantum fluctuations in our formalism. 
The origin of the thermal fluctuation in our formalism is quite different from
that in the phenomenological theory of irreversible processes which is described by the classical Langevin equation with the temperature dependent random noise. 
Instead of the coordinates $x$ and ${\tilde x}$, we introduce new coordinates 
for harmonic oscillator as
\be
&&X(t)=\sqrt{(1+n)}x(t)-\sqrt{n}\tilde{x}(t) \,, \\
&&\tilde{X}(t)=\sqrt{(1+n)}\tilde{x}(t)-\sqrt{n}x(t) \,, \\
&&n=\frac{1}{e^{\beta\hbar\omega}-1} \,,
\ee 
and suppose that the stochastic equations for $X(t)$ and $\tilde{X}(t)$ are given by
\be
dX(t)&=&-\omega X(t) dt \nn \\
&&+\sqrt{\frac{\hbar(1+n)}{m}}dW-\sqrt{\frac{\hbar n}{m}}d\tilde{W} \,, \\
d\tilde{X}(t)&=&\omega \tilde{X}(t) dt \nn \\
&&+\sqrt{\frac{\hbar(1+n)}{m}}d\tilde{W}-
\sqrt{\frac{\hbar n}{m}}dW \,. 
\ee
Here the drift terms have no 
temperature dependence,
 but the random noise terms depend on both the temperature and the Planck constant $\hbar$.
 In the zero temperature limit $\beta\to\infty$, 
the random noise terms approach  standard Wiener processes
 with coefficient $\sqrt{\frac{\hbar}{m}}$.
 In high temperature limit, the $\hbar$ dependence of the random noise terms vanishes,
 as it is expected that the system becomes classical, and the random noise becomes 
$
\sqrt{2kT/(\omega m)}(dW-d\tilde{W})/\sqrt{2}
$\,.
This way the stochastic equation for $X(t)$ becomes
 the classical Langevin equation.

 Our future task is to apply our formalism
 to various cases, such as system with the anharmonic oscillator potential,
  and to extend formalism to quantum open systems \cite{Tbath,Arimitsu2}.


\section*{Acknowledgments}
One of the author (K.~K.) would like to thank Mr.~K.~Honda
for a fruitful discussion.





\bibliographystyle{model1a-num-names}
\bibliography{<your-bib-database>}







\end{document}